\pdfoutput=1
\documentclass[11pt,twoside]{article}
\usepackage{arxiv}
\usepackage{multirow}
\usepackage{amssymb}
\usepackage[utf8]{inputenc}
\usepackage[T1]{fontenc}
\usepackage{lmodern}
\usepackage{graphicx}
\usepackage[figurename=Fig.,labelfont=bf,labelsep=period]{caption}
\usepackage{subcaption}
\usepackage{amsmath}
\usepackage{xurl}
\usepackage{newtxtext,newtxmath}
\usepackage[colorlinks=true,citecolor=red,linkcolor=black]{hyperref}

\usepackage{apacite}
\def\BibTeX{{\rm B\kern-.05em{\sc i\kern-.025em b}\kern-.08em
    T\kern-.1667em\lower.7ex\hbox{E}\kern-.125emX}}

\begin{document}
\title{Moodle Usability Assessment Methodology using the Universal Design for Learning perspective}

\newcommand{\shorttitle}{Moodle Usability Assessment Methodology}

\author{Rosana Montes$^1$ \And 
    Liliana Herrera$^2$ \And 
    Emilio Crisol$^3$ \And \\
\\$^1$ Department of Software Engineering and\\
\textit{the Andalusian Research Institute in Data Science and Computational Intelligence (DaSCI),}\\
\textit{University of Granada, Granada, Spain.}\\
\textit{E-mail: rosana@ugr.es}\\
\\$^2$ University of Atlántico, Barranquilla, Colombia\\
\textit{E-mail: rosana@ugr.es}\\
\\$^2$ Department of Didactics and School Organization.\\ University of Granada, Granada, Spain\\
\textit{E-mail: ecrisol@ugr.es}\\
}

\date{\today}

\maketitle

This work has been published in the Turkish Online Journal of Distance Education.
Cite as:  Herrera, L., Crisol, E., Montes, R. (2025). MOODLE USABILITY ASSESSMENT METHODOLOGY USING THE UNIVERSAL DESIGN FOR LEARNING PERSPECTIVE. Turkish Online Journal of Distance Education, 26(3), 238-255. https://doi.org/10.17718/tojde.1510242

\begin{abstract}
The application of the Universal Design for Learning framework favors the creation of virtual educational environments for all. It requires developing accessible content, having a usable platform, and the use of flexible didactics and evaluations that promote constant student motivation. The present study aims to design a methodology to evaluate the usability of the Moodle platform based on the principles of Universal Design for Learning, recognizing the importance of accessibility, usability and the availability of Assistive Technologies. We developed and applied a methodology to assess the usability level of Moodle platforms, taking into consideration that they integrate Assistive Technologies or are used for MOOC contexts. We provide the results of a use case that assesses two instances for the respective Moodle v.2.x and v.3.x family versions. We employed the framework of mixed design research in order to assess a MOOC-type educational program devised under the principles of Universal Design for Learning. As a result of the assessment of Moodle v.2.x and v.3.x, we conclude that the platforms must improve some key elements (e.g. contrasting colors, incorporation of alternative text and links) in order to comply with international accessibility standards. With respect to usability, we can confirm that the principles and guidelines of Universal Design for Learning are applicable to MOOC-type Virtual Learning Environments, are positively valued by students, and have a positive impact on certification rates.
\end{abstract}





\section{Introduction}\label{sec-1}

It is essential to consider inclusive Virtual Education as a high-impact strategy to improve the coverage, relevance and quality of education at all levels and types of training. Virtual Learning Environments (VLEs) are web-based systems that facilitate online communication, collaborative work, the sharing of various types of resources or educational materials, assessment and student monitoring~\cite{Cassidy2016}. Designing VLEs for all means recognizing different populations, including students with functional diversity or disability, so not only the characteristics of the platform must be kept in mind, but also the particularities of the pedagogical materials, which can become barriers to learning if not properly designed~\cite{Vilaverde2020,Fermin2019,Crisol2019}.

When aiming for quality and equal opportunities, two essential elements that emerge are inclusive education an digital literacy. UNESCO~\cite{Unesco} highlights that teachers must recognize the appropriate use of technologies in the harmonization of the needs and particularities of each student. It is important to formulate and implement a wide range of learning strategies that respond precisely to the diversity of students. One of the paradigms that is gaining more strength every day, showing successful experiences at different educational levels, is the Universal Design for Learning (UDL). According to the Center for Applied Special Technology (CAST)~\cite{CAST2011}, UDL focuses on the design of flexible and versatile educational curricula, with diverse materials and media so that everyone has access to learning. UDL takes advantage of the great benefits of technological media, such as: versatility, transformation capacity and generation of connections. E-learning offers a wide range of possibilities to promote Education for All, which is why it is widely used. According to the World Health Organization (WHO)~\cite{Who}, Assistive Technologies (AT) are systems and services related to software that provide help and support, allowing people to lead a productive and independent life and to be part of the social, labor and educational environment.

Accessibility and usability promote the possibility for users to use learning platforms and educational digital content according to their needs, participating equitably in digital educational environments. Designing virtual learning environments for all implies recognizing different populations, including students with functional diversity or disabilities, so it is necessary to keep in mind not only the characteristics of the platform, but also the particularities of the pedagogical materials. 

This research provides a methodology for the evaluation of usability in LMS (Learning Management System), Moodle v.2.x and v.3.x environments, which could well be transferable to other LMS environments, since it seeks to improve the VLE, in terms of usability and accessibility for all. It is important to mention that the v.2.x platform instance used corresponds to that used by the University of Granada in Spain\footnote{PRADO \url{https://prado.ugr.es}}, while the v.3.x instance is the one used at the Universidad del Atlántico in Colombia\footnote{SICVI 567 \url{https://sicvi567.uniatlantico.edu.co/}}, for the academic year 2019/20. As part of the evaluation, the design and use of a MOOC training action (implemented in the Moodle v.3.x platform) has been provided, incorporating the opinion of users in their different roles and recognizing the importance of education for all and universal learning. This experience opens the way to verifying the applicability of UDL principles as a key element to offer virtual educational programs for all, recognizing the importance of accessibility and usability of AT-supported environments.

The article is organized as follows. Inspired by the findings of CAST~\cite{CAST2011}, Section~\ref{sec-2} presents a literature review, starting from the importance of design for all people in the educational and technological field, and more specifically in online learning platforms. Section~\ref{sec-3} presents the methodology used for the study, explaining the mixed design and the instruments used for evaluating the Moodle platform in its two versions from the perspective of the roles of student, teacher, administrator and usability experts. Section~\ref{sec-4} describes the results, organized into three segments: applicability of AT in Moodle, usability of the Moodle platforms v.2.x and v.3.x, and evaluation of the MOOC course. Section~\ref{sec-5} is the discussion, which contrasts the findings with the theoretical corpus. Finally, Section~\ref{sec-6} presents the conclusions, perspectives and future projects to consolidate the proposed methodology.

\section{Literature review}\label{sec-2}

\subsection{Education for all}

Education for all, or Inclusive Education, is an educational approach that values diversity as an element that enriches the teaching-learning process and therefore the development of the human being. It is associated with values such as participation, equity, equality and justice, which transcend the attitudes of teachers, students and the community in general~\cite{Haug2017,Robinson2020}. Inclusive Education promotes learning for all, including students with special educational needs, by enabling equal opportunities, removing barriers, and through presence, participation and progress in physical and virtual contexts. Students with special educational needs require additional or different support to what is normally available in the teaching-learning process for other individuals of similar age~\cite{Florian2019}. 

Authors such as Kent et al.~\cite{Kent2018}, Rogers-Shaw et al.~\cite{Rogers2018} and Oswald et al.~\cite{Oswald2018}, point out the importance of the universal design approach in e-learning, arguing that it provides adaptations that favor the academic performance of students, especially those with special educational needs. The UDL framework invites the design of learning environments commensurate with the increasing diversity of learners encountered in online courses~\cite{Rogers2018}.

\begin{figure}
\centering
\includegraphics[width=.8\linewidth]{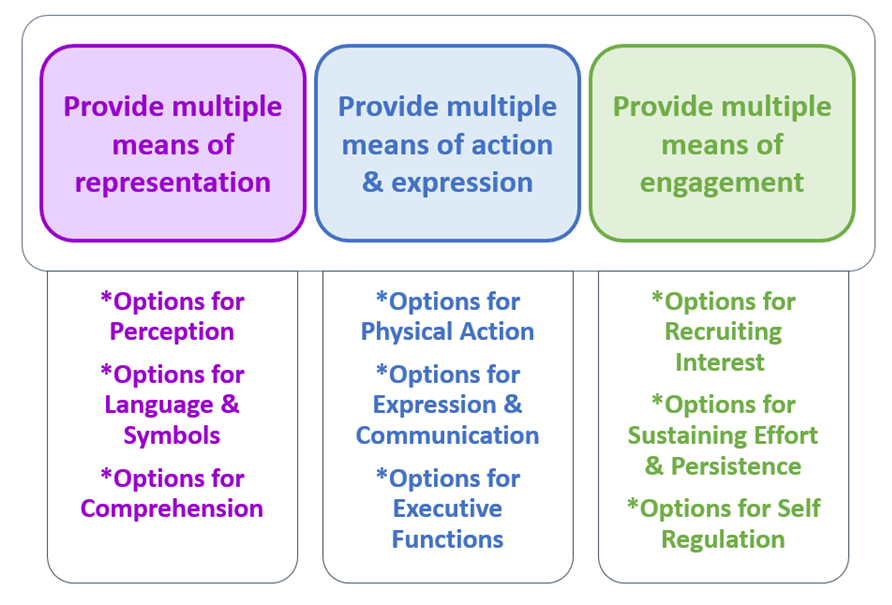}
\caption{CAST UDL Principles and Guidelines.}
\label{fig:uno}
\end{figure}

The three principles of UDL consist of promoting multiple forms of representation, expression and involvement \cite{CAST2011}. Figure ~\ref{fig:uno} graphically presents the three UDL principles and their guidelines. 

UDL is an approach that addresses the challenge of making all students expert learners and is widely used at various levels, including university~\cite{Dickinson2020,Frumos2020,Griful2020}. When choosing UDL implementation strategies, it is recommended that its principles be introduced gradually, starting by replacing undervalued activities to keep students and teachers excited about learning~\cite{Rao2015}.

\subsection{Technologies for everyone}

Technological accessibility and usability is vital for an information and knowledge society. Web accessibility refers to the possibility for all users, including those with functional diversity, to perceive, understand, navigate and interact with web contents, thanks to the fact that they follow design guidelines that allow access to people with a wide range of hearing, movement, visual and cognitive abilities~\cite{w3cac}.

Usability is a broader concept than accessibility; therefore, for a resource or page to be usable, it must be accessible in the first instance. However, despite the fact that accessibility, usability and integration of technologies are seen as reasonable practices, they are not always used in online courses~\cite{Edwards2018}. Assessing the accessibility of digital educational resources and online learning platforms is very important to ensure education for all, especially for students with disabilities~\cite{Alahmadi2017,Edwards2019}.

Assistive Technologies –-which for this case are technologies used to favor accessibility-– designate the systems and services related to programs for assistance, which allow people to lead a productive and independent life and be part of the social, labor and educational spheres~\cite{Who}. ATs enable access to educational information and content and to learning platforms themselves~\cite{Alahmadi2017}.

Research, such as that undertaken by Wu~\cite{Wu2015}, Mironova et al.~\cite{Mironova2016}, Fidalgo \& Thormann~\cite{Fidalgo2017} and Coleman \& Berge~\cite{Coleman2018}, shows that the accessibility needs, requirements and preferences of participants in virtual courses hosted on LMS platforms are diverse and that simplifying navigation is of great help~\cite{Rao2015}. In addition, Chatterjee et al.~\cite{Chatterjee2017} and Houston~\cite{Houston2018} describe the application of UDL principles in VLE, specifying that they offer alternatives for information presentation, interaction, expression and involvement with online learning and incorporating emerging technologies such as immersive education and simulators. Rogers-Shaw et al.~\cite{Rogers2018} and Ciasullo~\cite{Ciasullo2018} state that Inclusive Virtual Education is possible if it is designed with a methodology that involves knowing the students, developing accessible content, having a usable platform, and includes flexible didactic and evaluations that promote constant student motivation.

\subsection{Accessibility and usability evaluation of Moodle LMS}

There are several standards and norms to consider in the design of a web page, which seek to make it more accessible. Yet although there are specific recommendations, this does not guarantee that web designers take them into account~\cite{Armano2018}. To evaluate the accessibility of web content, two methods are used, one automatic and the other manual, which complement each other. The Moodle platform has been the subject of analysis in different distribution versions, regarding its accessibility~\cite{Moreno2011,Calvo2014}. Since then, the scope of measurement has been extended from accessibility to a broader concept such as usability.

In a usability test, the user is presented with the product and asked to use it intuitively or is asked what he/she thinks it can be used for~\cite{Krug2000}. No single method should be used, but rather they are combined for best results. The Moodle platform has been evaluated using usability principles in several of its versions. Kakasevski et al.~\cite{Kakasevski2008} evaluated and compared the user experience when using different Moodle modules, giving recommendations for teachers and students that make it possible to improve authentication tasks, learning resources, assignments and communication. In 2013, Daneshmandnia~\cite{Daneshmandnia2013} carried out a usability evaluation on Moodle v.2.4. through expert inspection, revealing positive responses regarding the possibilities of this LMS. This result coincides with that found by Kipkurui et al.~\cite{Kipkurui2014}, who identified that the learnability of the platform affects its usability. Orfanou et al.~\cite{Orfanou2015} propose the use of the SUS scale~\cite{brooke1996} to evaluate the perception of usability of Moodle platform users, finding that it is at a satisfactory level according to the students who participated.

\section{Methodology}\label{sec-3}

This study has been developed using a mixed design, which allows a series of systematic, empirical and critical processes that involve collecting qualitative and quantitative data. It was designed as a multi-method study adapted to our needs, contexts, circumstances and resources, in order to achieve a better understanding of the analyzed phenomenon~\cite{HdezSampieri,McKim2017}.

The objective is to analyze the viability of Moodle v.2.x and v.3.x for applying the UDL, based on an evaluation of its accessibility and usability. The objective itself is framed within the design of the evaluation of an educational program. The responses of this evaluation may or may not directly judge the value of an object, which in this case corresponds to the usability of two Moodle platforms. One platform is at the University of Atlántico (UA) in Colombia, using version 3.3, and the other is at the University of Granada (UGR) in Spain, with version 2.7. Both platforms reflect the two production environments that the Moodle.org community makes available to centers and institutions, generally referred to as Moodle v.2.x and v.3.x.

\subsection{Sample Sets}

Given the mixed characteristics of the study, it was necessary to have different samples to collect the relevant qualitative and quantitative data. In particular, the users considered are all strata of the same universe, i.e. with experience in the use of the LMS Moodle platform, since these selections reflect different levels of interaction with the platform. Due to the instruments used, it was necessary to define different sampling techniques, also taking into account that one of the characteristics of mixed methods is the possibility of combining different sampling techniques that make it possible to solve the problem statement~\cite{McKim2017}.

\begin{enumerate}
    \item \textbf{First sample}. This was composed of 9 people belonging to the coordinating and technical team of the Moodle platform in the two universities studied. All the participants work in the IT and e-learning areas, and most of them have between 6 and 20 years of experience.
    
    \item \textbf{Second sample}. This comprised a type of volunteer participant, i.e. individuals linked to undergraduate studies at the two institutions, who shared their experiences in the Usability Test (UT) of the Moodle platforms. There were 213 voluntary participants, 126 people from the University of Atlántico and 87 from the University of Granada.

    \item \textbf{Third sample}. This was formed by usability experts from both institutions – professionals in IT who are familiar with the platforms but who are not currently directly linked to the departments that administer them. There were 3 experts from the University of Atlántico and 4 from the University of Granada, for a total of 7 people.

    \item \textbf{Fourth sample}. This was made up of Moodle platform users in their role as student teachers or technical team members. A total of 22 people participated in these groups, which will be explained below.
    
    \item \textbf{Fifth sample}. This consisted of 531 participants on the MOOC course held at the University of Atlántico, who answered the second and fourth questionnaires to evaluate this educational program. Of this sample, 21 participants stated that they had some type of functional diversity; 50 had entered the educational system late; 30 had difficulties in mathematics; 6 had difficulties in reading and writing; 7 had attention deficit disorder with or without hyperactivity; and 1 had autism spectrum disorder.

\end{enumerate}

\subsection{Instruments}

Considering the mixed design of the present study, different types of data collection instruments have been designed, which are presented in Table~\ref{table:one} and detailed below. Template files and also data from our case study are publicly available as a GitHub repository, at \url{https://github.com/ari-dasci/OD-Moodle-Usability-Assessment}.

\begin{table}
\caption{List of instruments used in this research and type of data}
\label{table:one}
\centering
\small
\renewcommand{\arraystretch}{1.25}
\begin{tabular}{l l}
\hline\hline
\multicolumn{1}{c}{Instrument} &
\multicolumn{1}{c}{Type of data} \\
\hline
An AT Matrix & Qualitative \\
Report from Wave accessibility evaluation tool & Quantitative  \\
Usability Testing & Mixed \\
Focus groups & Qualitative \\
Seven closed questionnaires & Quantitative \\
\hline\hline
\end{tabular}
\normalsize
\end{table}

\subsubsection{Assistive Technology Matrix}

The purpose of our AT matrix\footnote{AT matrix \url{https://github.com/ari-dasci/OD-Moodle-Usability-Assessment/blob/main/es/Anexo-para-cuestionarios-TA-matrix.pdf}} is to systematize in an organized way the evidence found about the search for technology that is free and online, and likely to be used in VLEs by teachers and students, with the aim of enhancing communication, reading, writing, mathematics exercises, accessibility and organization tasks. The information collected in the matrix is qualitative and its results became an annex to one of the questionnaires addressed to the technical team that manages the Moodle platform. The 24 selected AT instances are classified into technologies for communication, accessibility, reading, writing, mathematics and time organization. 

\subsubsection{Web Accessibility Evaluation}

For the automatic evaluation of accessibility, the Wave\footnote{Wave Web Accessibility Evaluation Tool \url{https://wave.webaim.org/}} tool was selected. The tool loads the web page, and detects and marks accessibility errors on it. When these errors are clicked upon, a detailed report is given, thus facilitating understanding. For its interpretation, we used the Website Accessibility Conformance Evaluation Methodology (WCAG-EM) 1.0\footnote{WCAG-EM 1.0 \url{https://www.w3.org/TR/WCAG-EM/}}, a methodology that determines the accessibility level of a web page according to Web Content Accessibility Guidelines 2.0\footnote{WCAG 2.0 \url{https://www.w3.org/TR/WCAG20/}}. This is a flexible support resource that does not define additional requirements, nor does it determine a specific tool for its application, and it allows self-evaluation and evaluation by third parties~\cite{w3cwcag}.

In the following GitHub resource\footnote{Web Accsibility report \url{https://github.com/ari-dasci/OD-Moodle-Usability-Assessment/blob/34e40cbce1da37e5fc98495f8aa299e9a5e3068e/es/Informe-AccesibilidadWeb-Wave.pdf}} you can read our analysis for PRADO and SICVI567 that encompasses both Wave accessibility reports.

\subsubsection{Usability Testing}

 For the evaluation, the MOOC course title \textit{Inclusive educational contexts: design for all people} was hosted on both LMS platforms, and the 12 review pages of the course were defined and organized according to the tasks to be used in the usability test\footnote{UT task \url{https://github.com/ari-dasci/OD-Moodle-Usability-Assessment/blob/main/es/tareas_testU_moodle.pdf}}.

\subsubsection{Focus Group for Moodle users}

Focus groups, also called discussion groups, are sessions in which topics are discussed in depth. These consisted of meetings of people where meanings were constructed in a collaborative way, guided by a moderator. In this case, positive and negative aspects were discussed, as well as suggestions for improvement of the UA's Moodle platform. The focus group was structured by means of two techniques proposed by Design Thinking, the Journey Map and the Feedback-Capture-Grid\footnote{Focus Group responses \url{https://github.com/ari-dasci/OD-Moodle-Usability-Assessment/blob/main/es/Informe-GruposFocales-UA-respuestas.pdf}}. These techniques are methods that make it possible to analyze and discover people's needs, under the paradigm of universal design, and which therefore may help to improve the design of a MOOC course. We had the participation of 22 people from the UA, belonging to the fourth sample described above. More information is given in our report\footnote{Focus Group at UA \url{https://github.com/ari-dasci/OD-Moodle-Usability-Assessment/blob/main/es/Informe-GruposFocales.pdf}}.

\subsubsection{Questionnaires}

In the search for a reliable, valid and objective process, three main stages were carried out: design, validation and piloting. These stages led to the final version of the questionnaires. The validation process of seven questionnaires was carried out using the online modified Delphi Method and with the participation of ten expert judges. All of them are professionals in the field of education and engineering with extensive experience in the field of virtual education and inclusive education. The experts' evaluations were analyzed quantitatively and qualitatively. The pertinent qualitative modifications were made, and in order to evaluate the internal consistency of the questionnaires, a pilot test was applied. We then carried out Cronbach's Alpha on its results. The Cronbach's Alpha index ranges from 0 (0\%) to 1 (100\%), where the minimum expected was 0.8 (80\%), and it showed the degree of concordance of the questionnaire responses. The values obtained in the questionnaires were higher than 80\%, thus demonstrating their internal consistency~\cite{Bujang2018}

The total set of questionnaires is listed in Table~\ref{table:two}, and their template can be downloaded at the GitHub repository\footnote{Instruments \url{https://github.com/ari-dasci/OD-Moodle-Usability-Assessment/blob/main/instruments.md}}. It is important to mention that a five-term Likert scale was used to answer the questionnaires: \textit{strongly agree, partially agree, indifferent, partially disagree, strongly disagree}.

\begin{table}
\caption{Target audience and purpose of each questionnaire}
\label{table:two}
\centering
\small
\begin{tabular}{|p{4cm}|p{9cm}|} 
\hline\hline
\multicolumn{1}{c}{Instrument} &
\multicolumn{1}{c}{Purpose} \\
\hline
Q1: Managers of Moodle platform & To determine the opinion regarding the attention to diversity in the VLE and the integration of AT tools in Moodle and how these can favor its usability. \\
Q1: Technical administrators & Determine the opinion regarding the integration of AT tools to the Moodle platform and how these can favor its usability.  \\
Q2: Moodle user teachers & To determine the opinion about the usability of the Moodle platform of users with experience in the teaching role. \\
Q2: Moodle user students & Determine the opinion about the usability of the Moodle Platform of users with a student role. \\
Q3: Usability experts & Determine the opinion regarding the usability of the Moodle Platform. Evaluate 10 usability dimensions or attributes: Web Accessibility, Identity, Navigation, Efficiency, Effectiveness, Help, Content, Easy to Remember, Satisfaction, and Resource Accessibility. \\
Q4: MOOC participants with experience in teaching & Determine the opinion regarding the application of UDL principles in their pedagogical practice in virtual learning environments.\\
Q4: MOOC Moodle participants & Determine the opinion regarding the MOOC course Inclusive Educational Contexts: Design for All, which was designed under the principles of UDL.\\
\hline\hline
\end{tabular}
\normalsize
\end{table}

The instruments used in the research enabled a broad analysis of the usability of the Moodle platform and how the application of the UDL in VLE can meet the needs of users of this type of technological tool. It was important to discover the usability of the Moodle platform in its v.2.x and v.3.x versions from the perspective of users with teaching, student and administrative roles, as well as to carry out usability evaluations using automatic tools and taking into consideration the opinion of experts. In this study, the Wave automatic accessibility evaluation tool and the Q3 questionnaire were used to obtain an evaluation on accessibility and usability, respectively.

It was also important to learn the opinion of managers and technical staff who administer Moodle platforms about online AT tools that could be integrated into the LMS platform to improve learning conditions and promote education for all. The Q1 instrument was used to collect information from both groups of users. Furthermore, we needed to find out the opinion of the MOOC course participants, before and after, in order to learn about their experience on the Moodle platform and to determine if the MOOC, designed under the principles of the UDL, fully met its purposes and was adapted to the needs of the participants. 
All the instruments used shed light on the usability not only of the Moodle platform in two of its versions, but also the course that was hosted on it and designed under UDL principles, with an emphasis on the Q2 and Q4 questionnaires. The quantitative data were collected with the Wave accessibility assessment tool, the questionnaires and the UT, while the qualitative data were processed using NVivo 11\footnote{NVivo \url{https://qsrinternational.com/nvivo}}.

\subsection{Assessment under the Computing with Words Paradigm}\label{sec-LDM}

Computing with Words (CW) is a methodology that uses human perceptions and opinions to reason and calculate instead of numerical measures. CW has the ability to optimize the approach experts use to calculate assessments of decision-making problems. To complement the use of CW, it is necessary to use linguistic variables. A linguistic assessment was carried out to provide an overall estimate of all the results obtained by applying the instruments, based on fuzzy set theory~\cite{ZADEH1975}, which is widely used for the processing of qualitative data, based on the methodology of computing with words~\cite{Mendel2010}. Thanks to this methodology, algorithms can be developed that operate with data expressed by means of linguistic variables, although this research focuses only on the assessment and not so much on their processing.

Fuzzy linguistic valuation makes it possible to represent qualitative aspects, based on linguistic variables, which do not have numerical values but words expressed in natural or artificial language~\cite{ZADEH1975}. Each linguistic value is characterized by a label and a meaning, where the label is a word that belongs to a set of linguistic terms, and the meaning is what it refers to and can be interpreted depending on the context. 

In this research we use a term set $S$ of $n=6$ that gives domain to the usability linguistic variable, with $S= \{worst, poor, good, very\_good, excellent, unmatched\}$. In this way $s_i \in S,\; i=(1,\dots,n)$ is a valid output of the Brooke \textit{System Usability Scale} (SUS) questionnaire~\cite{brooke1996}. For instance, $usability=s_2$ (which is \textit{poor}) is a better approximation of the computed SUS score in the range $[0,100]$. This questionnaire is simple, since it has an input of a 10-item scale that allows a global evaluation of usability. Linguistic rating is adopted in this study because it allows the comparison of platforms in terms of usability values expressed as human perceptions and opinions, which are more understandable than numerical measures.

\section{Results}\label{sec-4}

The results are presented in three sections and give an account of the analysis carried out from the application of the instruments. They allow us to recognize the perspectives of different users of the Moodle platform, as well as the results of the accessibility and usability tests. 

\subsection{Applicability of Assistive Technologies in Moodle}

As explained in Section 3.2.1, we created the AT matrix resource to collect 24 solutions that cover different potential needs for users. The purpose was to find online AT tools that could be made available to Moodle platform users, in order to improve the Moodle platform configuration for usability. Among the AT tools selected are interpretation services in Spanish Sign Language and Colombian Sign Language, optical character recognition (OCR) tools, dictation tools, text to audio converters and vice versa, grammar checkers, dictionaries, and mathematical equation editors, among others.

The analysis of the results of the Q1 questionnaires showed that only one coordinator, corresponding to 20\%, agreed that teachers requested support in the use of ATs to assist students with special educational needs. However, the lack of appropriate use of these tools by the coordinators of the virtual education services was evident. Fifty percent of the members of the technical team recommended the integration of six specific online ATs, such as the sign language interpretation service\footnote{SVIsual software \url{http://www.svisual.org}}, the relay center\footnote{Centro de Relevo services (Colombia) \url{https://www.centroderelevo.gov.co/}}, the RAE dictionary\footnote{Spanish Royal Language Academy's Dictionary \url{http://dle.rae.es/}}, Tomatoes\footnote{Tomatoes software \url{http://www.tomato.es/}}, and the two online equation editors\footnote{LaTeX equation editor \url{https://www.codecogs.com/latex/eqneditor.php}}. Similarly, 75\% agreed on recommending the Checklist for Moodle. All respondents stated that the incorporation of ATs into the Moodle platform favors inclusion and stated that they require training to learn how to use them.

\subsection{Usability assessment}

Usability refers to the ease of use of the Moodle LMS platform. To be usable, it must be accessible in the first place, so it is pertinent to evaluate the accessibility of the v.2.x and v.3.x versions, as we describe in the following five subsections.

\subsubsection{Accessibility evaluation results}\label{aspecto-1}

Wave was used as an online service for web site accessibility evaluation and was applied to Moodle v.2.x and v.3.x instances. The tool was applied to the five groups of tasks and pages to which the user has access.

Table 3 synthesizes the metrics obtained for the v.2.x instance in terms of errors, alerts, features and accessibility according to tasks. The accessibility column indicates the global labels obtained according to the level of compliance with the standards proposed by the W3C. The usability column, added by us, allows us to provide a linguistic interpretation of usability according to all the results. This assessment is given by the researcher when reviewing the results of the automatic reports. For the case analyzed, it has a negative impact on usability, since it is not accessible to people with functional diversity, and is not very usable in general. It was observed that the pages with the highest number of errors are those related to grade tracking, edit profile and glossary activity, with 96, 74 and 72 errors respectively. 


\begin{table}[h!]
\caption{Report by Wave in Moodle v.2.x}
\label{table:three}
\centering
\small  \scriptsize 
\renewcommand{\arraystretch}{1.25}
\begin{tabular}{l l l l l l l l}
\hline\hline
Task type &
Task & 
Page & 
Errors & 
Alerts & 
Features & 
Accessibility &
Usability \\
\hline
\multirow{3}{4em}{Platform Login} & \mbox{Moodle login} & login-dialog & 16 & 13  & 8 & A & Poor\\
& \mbox{Find Course} & home-page & 49 & 22 & 33 & A & Poor  \\
& \mbox{Course access}& course-page & 46  & 121 & 156 & A & Poor\\
\hline
\mbox{Account settings} & \mbox{Change preferences} & edit-profile & 74  & 84 & 161 & A & Poor \\
\hline
\multirow{2}{4em}{Course view} & \mbox{Content access} &  content-fetching & 39 & 29  & 93 & A & Poor\\
& \mbox{Access to news}& page-fetching & 39 & 33  & 101 & A & Poor\\
\hline
\multirow{3}{4em}{Communications} & \mbox{Forum access} & forum-main  & 39 & 31 & 94 & A & Poor\\
& \mbox{Send a message} & messaging  & 56 & 43 & 116 & A & Poor\\
& \mbox{Chat participation} & blogging  & 41 & 27 & 87 & A & Poor\\
\hline
\multirow{3}{4em}{Activities} & \mbox{Task access} & task-view  & 46 & 29 & 89 & A & Poor\\
& \mbox{Questionnaire access} & questionnaire-view  & 7 & 29 & 97 & A & Poor\\
& \mbox{Glossary access} & glossary-view  & 72 & 35 & 93 & A & Poor\\
& \mbox{Rating follow-up} & rating-view  & 96 & 48 & 101 & A & Poor\\
\hline
\multicolumn{3}{r}{Total} & 620  & 544 & 1229 & A & Poor\\
\hline\hline
\end{tabular}
\normalsize
\end{table}

Table~\ref{table:four} shows the results of the Moodle v.3.x evaluation. It was observed that the pages where the highest number of errors, alerts and features were concentrated were the course home page with 103 incidents, profile editing in preferences with 71, and messaging with 49 errors reported. The errors detected by the tool indicated accessibility problems that need to be corrected, as they pose difficulties for the end user. 

\begin{table}[h!]
\caption{Report by Wave in Moodle v.3.x}
\label{table:four}
\centering
\small  \scriptsize 
\renewcommand{\arraystretch}{1.25}
\begin{tabular}{l l l l l l l l}
\hline\hline
Task type &
Task & 
Page & 
Errors & 
Alerts & 
Features & 
Accessibility &
Usability \\
\hline
\multirow{3}{4em}{Platform Login} & \mbox{Moodle login} & login-dialog & 6 & 5  & 4 & A & Poor\\
& \mbox{Find Course} & home-page & 31 & 28 & 29 & A & Poor  \\
& \mbox{Course access}& course-page  & 103 & 83 & 212 & A & Poor\\
\hline
\mbox{Account settings} & \mbox{Change preferences} & edit-profile & 71  & 31 & 60 & A & Poor \\
\hline
\multirow{2}{4em}{Course view} & \mbox{Content access} &  content-fetching & 10 & 24  & 21 & A & Poor\\
& \mbox{Access to news}& page-fetching & 10 & 30  & 33 & A & Poor\\
\hline
\multirow{3}{4em}{Communications} & \mbox{Forum access} & forum-main  & 9 & 23 & 19 & A & Poor\\
& \mbox{Send a message} & messaging  & 49 & 32 & 51 & A & Poor\\
& \mbox{Chat participation} & blogging  & 10 & 21 & 14 & A & Poor\\
\hline
\multirow{3}{4em}{Activities} & \mbox{Task access} & task-view  & 14 & 25 & 17 & A & Poor\\
& \mbox{Questionnaire access} & questionnaire-view  & 7 & 18 & 13 & A & Poor\\
& \mbox{Glossary access} & glossary-view  & 37 & 23 & 17 & A & Poor\\
& \mbox{Rating follow-up} & rating-view  & 12 & 27 & 23 & A & Poor\\
\hline
\multicolumn{3}{r}{Total} & 362  & 370 & 422 & A & Poor\\
\hline\hline
\end{tabular}

\normalsize
\end{table}

\subsubsection{Usability Testing}\label{aspecto-2}

To carry out this evaluation, a course was organized in Moodle v.2.x and v.3.x platforms and 28 tasks were defined for this test. The success of the completion of each task was recorded (if achieved/not achieved), as well as the time in seconds used and the emotion generated in the participant when he/she faced that task. This value is given by the linguistic variable $emotion=\{negative, neutral,positive\}$. A total of 213 users with the role of students participated, of whom 15 reported having special educational needs such as hearing, visual or organic disabilities and late entry into the educational system.

Table~\ref{table:five} presents a summary of the tasks that participants successfully completed in both versions of the Moodle platform and the average time it took them to complete them expressed in seconds. The emotion column shows the most predominant and estimated average results for each task. Lastly, the column labeled usability has been graded based on the results obtained and using the fuzzy linguistic approach explained in Section~\ref{sec-LDM}.

\begin{table}
\caption{Summary by Task Accomplishment at UGR and UA: Average Time of Duration (ATD) and the linguistic variable $emotion=$\{negative (-), neutral (0), positive (+)\}. }
\label{table:five}
\centering
\small \scriptsize 
\renewcommand{\arraystretch}{1.25}
\begin{tabular}{p{1.8cm}|p{2.6cm}|l|c|c|c|c|c|c|c|c}
\hline\hline
\multirow{2}{4em}{Task Type} &
\multirow{2}{4em}{Tasks} &
\multirow{2}{4em}{Estimated} &
\multicolumn{2}{|c|}{Success} & 
\multicolumn{2}{|c|}{ATD} & 
\multicolumn{2}{|c|}{Emotion} & 
\multicolumn{2}{|c}{Usability} \\
&  & & v2x & v3x & v2x & v3x & v2x & v3x & v2x & v3x \\
\hline\hline
\multirow{3}{4em}{Platform Login} & \mbox{\scriptsize 1. Moodle login} & 20-30 s & 100\% & 100\% & 27 s & 52 s & \scriptsize 0 & \scriptsize + & \scriptsize Unmatched & \scriptsize Unmatched \\
 & \mbox{\scriptsize 2. Find Course} & 20-30 s & 90\% & 100\% & 42 s & 49 s & \scriptsize 0 & \scriptsize + &\scriptsize Unmatched &\scriptsize Unmatched \\
 & \mbox{\scriptsize 3. Course access} & 10-20 s & 100\% & 98\% & 7 s & 4 s &\scriptsize 0 & \scriptsize + & \scriptsize Unmatched & \scriptsize Excellent \\
\hline
\multirow{3}{4em}{Technical Support} & \mbox{\scriptsize 4. Locate FAQ} & 40-50 s & 88\% & 55\% & 49 s & 60 s & \scriptsize 0 & \scriptsize - & \scriptsize Unmatched & \scriptsize Poor \\
& \mbox{\scriptsize 5. Contact form} & 60-90 s & 90\% & 25\% & 30 s & 174 s & \scriptsize 0 & \scriptsize - & \scriptsize Unmatched & \scriptsize The worst \\
& \mbox{\scriptsize 6. Change language} & 60-90 s & 93\% & 70\% & 21 s & 104 s & \scriptsize 0 & \scriptsize + & \scriptsize Unmatched & \scriptsize Very Good \\
\hline
\multirow{2}{4em}{Account settings} &
\mbox{\scriptsize 7. Set preferences} & 60-120 s & 90\% & 90\% & 72 s & 85 s & \scriptsize 0 & \scriptsize + & \scriptsize Unmatched & \scriptsize Excellent \\
& \mbox{\scriptsize 8. Upload image} & 60-120 s & 84\% & 78\% & 100 s & 139 s & \scriptsize 0 & \scriptsize + & \scriptsize Excellent & \scriptsize Very Good\\
\hline

\multirow{8}{4em}{Content access} & \mbox{\scriptsize 9. Access to news}  & 20-60 s & 77\% & 86\% & 63 s & 142 s & \scriptsize 0 & \scriptsize + & \scriptsize Excellent & \scriptsize Excellent\\

& \mbox{\scriptsize 10. Download a file} & 40-80 s & 98\% & 93\% & 29 s & 80 s & \scriptsize 0 & \scriptsize + & \scriptsize Unmatched  & \scriptsize  Excellent\\ 

& \mbox{\scriptsize 11. Directory download} & 40-80 s & 94\% & 78\% & 30 s & 95 s & \scriptsize 0 & \scriptsize + & \scriptsize Unmatched & \scriptsize Very Good\\

& \mbox{\scriptsize 12. Follow link} & 20-60 s & 96\% & 77\% & 26 s & 74 s & \scriptsize 0 & \scriptsize + & \scriptsize Unmatched & \scriptsize Very Good\\

& \mbox{\scriptsize 13. Embedded video} & 20-60 s & 96\% & 88\% & 35 s & 48 s & \scriptsize 0 & \scriptsize + & \scriptsize Unmatched  & \scriptsize Excellent\\
& \mbox{\scriptsize 14. Viewing a page}  & 40-80 s & 92\% & 90\% & 34 s & 66 s & \scriptsize 0 & \scriptsize + & \scriptsize Unmatched & \scriptsize Excellent\\
& \mbox{\scriptsize 15. Page Reading} & 40-80 s & 92\% & 87\% & 38 s & 81 s & \scriptsize 0 & \scriptsize 0 & \scriptsize Unmatched & \scriptsize Excellent\\
& \mbox{\scriptsize 16. Page Displaying}  & 40-80 s & 92\% & 85\% & 37 s & 72 s & \scriptsize 0 & \scriptsize + & \scriptsize Unmatched & \scriptsize Excellent\\

\hline
\multirow{2}{4em}{Communications} & \mbox{\scriptsize 17. Send a message}  & 50-90 s & 92\% & 81\% & 41 s & 116 s & \scriptsize 0 & \scriptsize + & \scriptsize Unmatched & \scriptsize Very Good\\
& \mbox{\scriptsize 18. Chat participation} & 60-120 s & 73\% & 79\% & 73 s & 107 s & \scriptsize 0 & \scriptsize 0 & \scriptsize Very Good & \scriptsize Very Good\\

\hline
\multirow{10}{4em}{Activities} &  \mbox{\scriptsize 19. Submit assignment}  & 60-120 s & 98\% & 76\% & 54 s & 129 s & \scriptsize 0 & \scriptsize + & \scriptsize Unmatched & \scriptsize Very Good\\
& \mbox{\scriptsize 20. Answer questionnaire}  & 60-120 s & 100\% & 87\% & 50 s & 116 s & \scriptsize 0 & \scriptsize + & \scriptsize Unmatched & \scriptsize Excellent\\
& \mbox{\scriptsize 21. Add to glossary} & 60-120 s & 99\% & 78\% & 120 s & 150 s & \scriptsize 0 & \scriptsize + & \scriptsize Unmatched & \scriptsize Very Good\\
& \mbox{\scriptsize 22. Set up groups}  & 60-120 s & 94\% & 11\% & 57 s & 108 s & \scriptsize 0 & \scriptsize - & \scriptsize Unmatched & \scriptsize The worst\\
& \mbox{\scriptsize 23. Adding to forum}  & 90-120 s & 87\% & 80\% & 93 s & 110 s & \scriptsize 0 & \scriptsize + & \scriptsize Excellent & \scriptsize Excellent\\
& \mbox{\scriptsize 24. Formatting text}  & 60-120 s & 79\% & 81\% & 97 s & 85 s & \scriptsize 0 & \scriptsize + & \scriptsize Excellent & \scriptsize Excellent\\
& \mbox{\scriptsize 25. Creating a link} & 60-120 s & 91\% & 69\% & 77 s & 99 s & \scriptsize 0 & \scriptsize + & \scriptsize Unmatched & \scriptsize Very Good\\
& \mbox{\scriptsize 26. Insert an image} & 60-120 s & 92\% & 80\% & 61 s & 79 s & \scriptsize 0 & \scriptsize + & \scriptsize Unmatched & \scriptsize Excellent\\
& \mbox{\scriptsize 27. Resize the editor} & 40-80 s & 72\% & 35\% & 59 s & 89 s & \scriptsize 0 & \scriptsize - & \scriptsize Very Good & Poor\\
& \mbox{\scriptsize 28. Rating follow-up} & 60-90 s & 92\% & 89\% & 34 s & 73 s & \scriptsize 0 & \scriptsize + & \scriptsize Unmatched & \scriptsize Excellent\\

\hline\hline
\end{tabular}
\normalsize
\end{table}

The block of tasks performed by between 100\% and 90\% are those related to the start of the platform. It should be highlighted that the student users of v.2.x mostly rated the completion of the tasks neutrally, while the users of v.3.x gave the positive rating to a greater extent to describe the emotion. The tasks of accessing the form with the platform's technical support and forming groups show significant differences between the users of the evaluated versions. We observed that the v.2.x users performed them more easily than the v.3.x users.

\subsubsection{Questionnaires Q2 and Q3}\label{aspecto-3}

With respect to the results of the Q2 questionnaires, addressed to teachers and student users of the platform, it was noted that both were in complete agreement that learning to use the platform and its new functions is simple. They also felt that they can explore the functionalities of the platform by trial and error. The sections of the platform were easily identified, expressing that they strongly (71\% and 53\%) and partially agree (41\%), respectively. Similarly, 46\% of the respondents (55 people) and 37\% (72 students) stated that the Moodle platform is similar to other common interfaces for teachers. Both types of users expressed that the platform is easy to use, even after not having used it, and they strongly agree (62\% and 55\%). A total of 77 teachers, representing 65\%, totally agreed that the functioning of Moodle is satisfactory, as did 117 students, corresponding to 60\%. Likewise, 93 teachers (78\%) and 129 students (66\%) stated that they would recommend the Moodle platform to people who might require an LMS system. It was observed that both teachers and students thought that the Moodle platform has an attractive interface (60\% and 46\%), its organization is clear (70\% and 52\%) and logical (65\% and 51\%) and that the environment is user-friendly (73\% and 60\%). The above shows a positive assessment of Moodle's usability.

The Q3 questionnaire aimed at usability experts was answered by engineers linked to the University of Atlántico and the University of Granada, but who are not responsible for the administration of Moodle platforms. The results showed that, according to the experts, both v.2.x and v.3.x had an adequate contrast between background and text, as well as a typography that facilitates reading. However, differences were observed between these platforms in terms of alternative text. The group of experts who evaluated the v.2.x. instance reported that they were indifferent to the use of the alternative text, while two experts from the v.3.x. evaluation group, representing 67\%, expressed a positive evaluation, and finally one expert (representing the remaining 33\%) expressed a negative evaluation. Regarding the hierarchical organization of the information, the v.2.x group stated that their opinion was indifferent, whereas for the v.3.x group, three experts (75\%) totally agreed that there is easy access to the information and partially agreed (50\%) that actions can be easily canceled.

The opinion of the expert group for v.3.x was partially in agreement on the question of whether the platform complies with navigability features. Opinions were more disparate with respect to the action cancellation feature. Sixty-seven percent (two experts) partially agreed that the help section is easy to locate, while each of the respondents expressed varying opinions about the help manual and contacting technical support. With respect to v.2.x, fifty percent (two experts) expressed that finding help is easy, seventy-five percent (three experts) expressed indifference to the manuals, and this same percentage expressed partial agreement that they have means of contacting the technical support team. Regarding the headers, it was observed that both v.2.x and v.3.x are clear and descriptive according to the experts, who were in full and partial agreement, respectively. A pop-up window is an element that appears on the screen, in this case, when accessing a section of the platform, and on this aspect, the v.2.x and v.3.x experts were in full agreement that the use of these is discreet.

Regarding the use of the highlighted sections, the majority of respondents in v.2.x were indifferent about this element of usability. For the v.3.x case, they strongly agreed. The evaluators stated that both v.2.x and v.3.x are pleasant and satisfactory to use, with data in the strongly agree column, as shown in Table~\ref{table:five}, and partially agree on the attractiveness of the v.3.x platform by 67\% (two experts). The Q3 questionnaire directed at experts was applied to evaluate the usability of both Moodle platforms. The results obtained in both indicated good usability, considering the linguistic evaluation given.

\subsubsection{Qualitative Data analyzed with NVivo}\label{aspecto-4}

In order to analyze the records obtained in the focus groups and the UT, as well as the presence of important elements, three categories of analysis identified through the use of NVivo software were chosen. The categories chosen by the researcher are the positive aspects, negative aspects and suggestions for improvement in the blocks of tasks faced by the users, as well as general comments on the Moodle platform. The most highly valued tasks (with a total of 12 references) are those related to the use of Activity resources. Users did not express comments or make positive references regarding the tasks of accessing technical support and user account management in either of the two sources analyzed. Likewise, no positive aspects were referenced in the UT regarding user account management, access to information and resources/content and general comments on the platform.

\subsubsection{Analysis based on computing with words methodology}\label{aspecto-5}

The linguistic variable usability allows us to represent the global view of the usability property, and we applied it to both platforms. It refers to an interpretation of the global results of the instruments that arises from the point of view of the research team. Table~\ref{table:six} shows the usability of both platforms in comparison. 

\begin{table}
\caption{Linguistic usability evaluation of Moodle v.2.x and v.3.x.}
\label{table:six}
\centering
\small  \scriptsize 
\renewcommand{\arraystretch}{1.25}
\begin{tabular}{lll}
\hline\hline
Instrument & Moodle v.2.x & Moodle v.3.x  \\
\hline
Wave accessibility evaluation & Poor & Poor \\
Usability Testing & Good & Good \\
Emotions in Usability Testing & Positive  & Positive \\
Q2: Moodle user teachers and students & - & Very good\\
Q3: Usability experts & Good & Good \\
Q4: MOOC participants with and without experience in teaching & - & Very good\\
\hline\hline
\end{tabular}
\normalsize
\end{table}

The conclusion reached is that the usability of the v.3.x platform was good, as was that of the v.2.x platform. However, these platforms require adjustments in their configuration to be made more accessible, and an intervention by the technical team that administers them, since the rating of poor was obtained by both. Likewise, a thorough review by the technical team is required to comply with accessibility standards. It should be made clear that questionnaires Q2 and Q4 were applied in the MOOC course developed only on the v.3.x platform; therefore there is no data for v.2.x.

\subsection{MOOC course evaluation}\label{aspecto-6}

The educational program proposed in this research is the MOOC \texttt{Inclusive Educational Contexts: Design For All} (MOOC-IEC). This course was designed based on the parameters of the UDL. The following are the results obtained at the end of the course, which reflect the opinion of the participants. First, the opinion of the participants with teaching experience in LMS platforms is shown, followed by the opinion of the participants on the design and experience after completing the course. Table 7 presents the results of the Q4 questionnaire in its two versions, one addressed to MOOC participants and the other addressed to participants with teaching experience in virtual educational environments. The first two columns of the table present the principles and guidelines of the UDL, followed by the Likert scale of the questionnaire.

\begin{table}
\caption{Questionnaire results according to UDL principles. Scale: Strongly disagree, Partially disagree, Indifferent, Partially agree, Strongly agree. Other acronyms: participant (P), participant with teaching experience (PTE)}
\label{table:seven}
\centering
\small  \scriptsize 
\renewcommand{\arraystretch}{1.25}
\begin{tabular}{l|l|c|c|c|c|c|c|c|c|c|c}
\hline\hline
\multirow{2}{4em}{UDL principle} &
\multirow{2}{4em}{Guideline} &
\multicolumn{2}{|c|}{SA} & 
\multicolumn{2}{|c|}{PA} & 
\multicolumn{2}{|c|}{I} & 
\multicolumn{2}{|c}{PD} &
\multicolumn{2}{|c}{SD} \\
&  & PTE & P & PTE & P & PTE & P & PTE & P & PTE & P\\
\hline\hline
\multirow{3}{4em}{I} 
& \mbox{\scriptsize Perception} & 65\% & 67\% & 24\% &19\% & 6\% & 10\% & 2\% & 1\% & 3\% & 3\%\\
& \mbox{\scriptsize Language and symbols}  & 79\% & 60\% & 11\% &17\% & 5\% & 20\% & 3\% & 2\% & 2\% & 1\%\\
& \mbox{\scriptsize Comprehension}  & 79\% & 87\% & 18\% & 12\% & 1\% & 1\% & 2\% & 0\% & 0\% & 0\%\\
\hline
\multirow{2}{4em}{II} 
& \mbox{\scriptsize Physical performance} & 73\% & 80\% & 20\% & 18\% & 3\% & 2\% & 4\% & 0\% & 0\% & 0\%\\
& \mbox{\scriptsize Execution}  & 79\% & 74\% & 18\% & 23\% & 1\% & 2\% & 1\% & 1\% & 1\% & 0\%\\
\hline
\multirow{3}{4em}{III} 
& \mbox{\scriptsize Pursuit of interests} & 75\% & 70\% & 21\% & 27\% & 1\% & 1\% & 3\% & 2\% & 0\% & 0\%\\
& \mbox{\scriptsize Effort and persistence}  & 87\% & 84\% & 9\% & 15\% & 2\% & 1\% & 1\% & 0\% & 2\% & 0\%\\
& \mbox{\scriptsize Self-regulation}  & 84\% & 79\% & 14\% & 15\% & 1\% & 6\% & 1\% & 0\% & 0\% & 0\%\\

\hline\hline
\end{tabular}
\normalsize
\end{table}

The results of the Q4 questionnaire show that participants with and without teaching experience expressed full agreement that the three UDL principles were accounted for, especially the options of effort and persistence and self-regulation, with values above 80\%. Regarding the fuzzy linguistic assessment of usability, the data showed a very good usability. In general, all rating results were concentrated on the \textit{Strongly Agree} and \textit{Partially Agree} scale values, both for participants with and without teaching experience.

\section{Discussion}\label{sec-5}

The VLE organizes didactic resources that focus on students by analyzing their characteristics and learning styles. The VLE favors autonomy and self-regulation, through the use of and access to the different resources available. VLEs are a necessity in all educational institutions and their accessibility is of the utmost importance for students, including those with disabilities (Alahmadi \& Drew, 2016; Kent et al., 2018; Crisol, Herrera \& Montes, 2019; Edwards, 2019; Vilaverde, 2020). 

One of the objectives of the study was to identify online Assistive Technologies (ATs) that can be integrated into Moodle in order to improve its usability. In this regard, we found that there are options that initially enhance accessibility. ATs facilitate both accessibility and the completion of tasks faced by users of virtual platforms, because they allow the use and manipulation of different formats that guarantee access to information and knowledge (Alahmadi \& Drew, 2016). There were different AT options according to their function, ranging from screen magnifiers, optical character recognition, text-to-audio converter and screen readers for people with visual functional diversity, task organizers for people with autism spectrum disorder, audio-to-text converter for people with motor functional diversity, and voice recorders for those with reading and writing difficulties. One of the great contributions of the online ATs selected in this study is their open and free use, since no investment is required by users to access them, and they thus become empowered with their use. Promoting accessibility in VLEs is necessary to ensure that all social groups have access to virtual education and is supported by a legal framework that is becoming increasingly important worldwide every day (Alahmadi \& Drew, 2016; Edwards \& Boyd, 2018; Edwards, 2019).

The members of the virtual education departments, both academic and technical managers, indicated that students with special educational needs participate in virtual classrooms. They also recognize a potential in AT tools for promoting inclusive education. However, they lacked mastery of them and were reserved when suggesting their integration in Moodle.

Our study recognizes accessibility as an important component of usability. No web page can be usable if it is not first accessible to users. Therefore, web accessibility and usability metrics were selected to be applied in the Moodle platform, using user-centered design techniques, and a theoretical review on usability and accessibility, and their application to the VLE, was undertaken. For accessibility, an automatic evaluation was completed using the free Wave Tool. This tool reviews and identifies accessibility difficulties that may occur on a web page. 

Both the v.2.x platform of the UGR and v.3.x of the UA mostly obtained results of single \texttt{A} labels, so the platform requires improvements to meet the diversity of its users and improve accessibility conditions according to the standards proposed by the W3C. Both platforms were linguistically evaluated with the adjective \textit{poor}, given the conditions they present and accessibility barriers for users with functional diversity.

As a complement to the selected usability metrics, in this research we applied a Usability Testing (UT) approach as a key method of inquiry into the Moodle platforms in the study.  Subjecting users to tasks that are carried out in Moodle, verifying whether they complete them, how long they do it and the emotions they feel when facing them, gave the opportunity to analyze the degree of usability of these systems. The evaluation was not limited to the verification of accessibility standards, because although these are both important and a legal requirement, it is essential to determine the factor of human interaction with the Moodle platform. Hence the relevance of the usability results, which in this case and according to the selected metrics, found that in the UT users took more time than estimated to complete the tasks of accessing the technical support form, updating the profile image and adding an entry to the glossary. Likewise, the tasks they had more difficulties completing were those of forming groups, expanding the editor's area and accessing the technical support form, leaving them with negative emotions.

With respect to the opinions of users of the Moodle platforms in the study, it is evident that although the experts state that the platforms meet the attributes of good usability, as expressed in the Q3 questionnaire, and that teachers and students who responded to Q2 reaffirm this by indicating that it is easy to learn to use, efficient, easy to remember, error tolerant, produces satisfaction and is attractive, there are suggestions for improvement made by users in the focus groups. 

The positive aspects are directly related to the activities that users can do on the course, while the negative aspects mainly concern the tasks associated with the start of the platform and access to content. Likewise, they also suggest improvements in the access to resources and contents. Overall, this allowed the researcher to give a linguistic valuation based on the adjectives proposed by the System Usability Scale, resulting in \textit{usability = poor} for v.2.x, and \textit{usability = good} for v.3.x.

Finally, a MOOC type educational program was designed, which presented an introductory module and the content organized into 3 modules, 42 educational resources in different complementary formats (audio, text, video) and 12 evaluation activities (6 mandatory and 6 optional). This was used as a context for data collection to discover the opinions of the participants with and without teaching experience, on the application of the UDL in the VLE. The products, which have been designed under the principles of equality, flexibility, simplicity and intuition, seek an environment in which everyone can participate, recognizing equal opportunities, respect for the rights of all and the participation of all students (CAST, 2011).

Participants rated the pilot experience positively, indicating, according to the answers obtained in the Q4 questionnaire addressed to participants, that the guidelines proposed by the UDL paradigm were satisfactorily met. Through the course design, platform configuration and content creation, multiple means of representation, forms of action and expression and forms of involvement with learning were provided. 

The design of this online training action brings benefits in quality inclusive virtual education, it improves accessibility, without the need for technical adjustments in the platform, and involves the participants in their learning. VLEs designed using the principles set forth in the UDL do not imply that the teacher develops a unique path for each student and their needs, but that the variability of this system is taken into account through the prediction of the possible ways in which students can learn in a flexible way (Dickinson \& Gronseth, 2020; Frumos, 2020; Griful-Freixenet, et al 2020).

This educational initiative is a successful complement to the academic experience provided by students, graduates, officials, teachers and external guests and contributes to the democratization of education. It was a space for continuous learning, where participants shared and discussed different opinions and experiences of what \textit{Design for All} entails in inclusive educational contexts. The statistics of the course show that a total of 520 people accessed the platform at least once and completed the activities that allowed 215 participants to be certified, corresponding to 27\% of the population. Offering learning paths adapted to the characteristics of the participants and maintaining adequate feedback in MOOC courses can reduce dropout rates~\cite{Borrella2022}. 

The fulfillment of the specific objectives made it possible to carry out the general objective of the study, which consisted of designing a methodology to evaluate the usability of the Moodle platform based on the principles of UDL. This methodology is transferable to other contexts and institutions that wish to consider virtual education for all. UDL applied to a VLE promotes learning opportunities, improving pedagogical practices for a great diversity of students, including those with and without functional diversity. This enhances student retention in online courses, thus tackling one of its great challenges, the high dropout rate.

\section{Conclusions}\label{sec-6}

Virtual Learning Environments have evolved over the years, and today web designers, content creators, and teachers are concerned about usability as a present feature, since, if they focus on the student as the protagonist in education, they must recognize the wide variety of learning styles and preferences for receiving information that they may have. 

It is important to offer virtual education for all, recognizing the diversity of students, according to their physical, psychological and social characteristics, as well as their previous experiences and interests. This is a challenge because it involves the consideration of technological and pedagogical aspects that guarantee quality and educational equity in learning environments mediated by technologies that are constantly evolving and are widely used by a diversity of users.

The proposal that VLEs should consider the accessibility of platforms and digital educational resources favors inclusion processes. The Universal Design for Learning (UDL) can be applied to virtual educational contexts, more specifically to training actions on the Moodle platform, since it makes content more flexible so that everyone can access information and learning more easily. 

The use of Assistive Technologies (AT) is a good option to improve accessibility to the information society and therefore to knowledge. These technologies allow access to the curriculum, activities, resources and communication for students with functional diversity. The design of VLEs for all should not be limited to the use of AT, but should incorporate pedagogical and instructional practices for students with and without disabilities, based on the principles of the UDL. ATs can be used by students with and without disabilities and support UDL.

It is important to mention that the staff in charge of supporting virtual teaching, both academics and technicians, indicate that students with special needs participate in virtual classrooms and recognize that these tools have the potential to enhance inclusive education. However, the diversity and disparity of technological solutions does not encourage a deep mastery of them, which is why they show reservations when suggesting their integration into Moodle. Therefore, it is essential that the technical/managerial and pedagogical teams of support centers constantly monitor and update new ATs for virtual education in universities, provided that they wish to provide an inclusive service.

\section{Acknowledgments}

This work was partly supported by the research project PID2020-119478GB-I00 of the Ministry of Economy, cofinanciated by FEDER (European Regional Development Fund - ERDF). Thanks are due to the 25 guest lecturers and the staff of the CEPRUD - Center for the Production of Resources for the Digital University (University of Granada) and the Virtual Education Project (University of Atlántico) for their valuable contributions in bringing the course to fruition.

\label{section:references}

\bibliographystyle{apacite} 
\bibliography{MoodleUsabilityAssessment}

\end{document}